\documentclass{aa} 
\usepackage{graphicx}
\usepackage{txfonts}
\usepackage{natbib}
\usepackage{url}
\bibpunct{(}{)}{;}{a}{}{,}

\newcommand{\be}{\begin{equation}}
\newcommand{\ee}{\end{equation}}

\def\apjl{ApJL}
\def\apjs{ApJS}

\newcommand{\Mpc}{$h^{-1}$\thinspace Mpc}

\newcommand{\vmh}{h^{-1}\mathrm{Mpc} }
\newcommand{\vgh}{h^{-1}\mathrm{Gpc} }

\begin{document}  

\title{Unusual  A2142 supercluster with a collapsing core: \\
distribution of light and mass
} 

\author {Maret~Einasto\inst{1} \and Mirt~Gramann\inst{1} 
\and Enn~Saar\inst{1,2}
\and Lauri Juhan~Liivam\"agi\inst{1,3} 
\and Elmo~Tempel\inst{1} 
\and Jukka~Nevalainen\inst{1}
\and  Pekka~Hein\"am\"aki\inst{4}
\and Changbom~Park\inst{5}
\and Jaan~Einasto\inst{1,3,6}
}

\institute{Tartu Observatory, Observatooriumi 1, 61602 T\~oravere, Estonia
\and
Estonian Academy of Sciences, Kohtu 6, 10130 Tallinn, Estonia
\and
Institute of Physics, University of Tartu, Ravila 14c, 51010 Tartu, Estonia
\and 
Tuorla Observatory, University of Turku, V\"ais\"al\"antie 20, Piikki\"o, Finland
\and 
School of Physics, Korea Institute for Advanced Study, 85 Hoegiro, Dong-Dae-Mun-Gu, Seoul 130-722, Korea
\and
ICRANet, Piazza della Repubblica 10, 65122 Pescara, Italy
}

\authorrunning{M. Einasto et al. }

\offprints{M. Einasto}

\date{ Received   / Accepted   }

\titlerunning{A2142 supercluster}

\abstract
{
Superclusters of galaxies can be used  to test cosmological models of the
formation and evolution of the largest structures in the cosmic web, and of galaxy and
cluster evolution in superclusters.
}
{
We study the distribution, masses, and dynamical properties of galaxy groups in the 
A2142 supercluster.
}
{
We analyse the global luminosity density distribution in the supercluster
and divide the supercluster into the high-density core and the low-density
outskirts regions.
We find galaxy groups and filaments in these regions,
calculate their masses and mass-to-light ratios and analyse their dynamical state
with 1D and 3D statistics.
We use the spherical collapse model to study the dynamical state
of the supercluster.
}
{
In the A2142 supercluster rich groups and clusters 
lie along an almost straight line forming the $50$~\Mpc\ long main body
of the supercluster. 
The A2142 supercluster has a very high density core surrounded
by lower-density outskirts.
The total estimated mass of the 
supercluster is $M_{\mathrm{est}} = 6.2\times~10^{15}M_\odot$.
More than a half of groups
with at least ten member galaxies in the supercluster lie in the 
high-density core of the supercluster, centred at the 
X-ray cluster \object{A2142}. Most of the groups in the
core region are multimodal.  In the outskirts of the
supercluster, the number of groups is larger than in the core,
and groups are poorer. 
The orientation of the axis of the cluster A2142 follows 
the orientations of its  X-ray substructures and radio halo, 
and is aligned along the supercluster axis. 
The high-density core of the supercluster
with the global density $D8 \geq 17$ and perhaps with $D8 \geq 13$
may have started to collapse.   
}
{ 
A2142 supercluster with collapsing core and straight body,
is an unusual object among superclusters. 
In the course of the future evolution,
the supercluster may split into several systems.      
}
%\end{abstract}


\maketitle

\section{Introduction} 
\label{sect:intro} 

Galaxy superclusters are the largest relatively isolated systems 
in the Universe \citep{1956VA......2.1584D, 1958ApJS....3..211A, 1978MNRAS.185..357J, 
1980Natur.283...47E, 1984MNRAS.206..529E,
1994MNRAS.269..301E, 1993ApJ...407..470Z}. 
Superclusters, with their high-density environment and complex inner structure, 
are excellent laboratories with which to study the properties and evolution
of galaxies and the groups of galaxies in them
\citep{2007A&A...464..815E, 2007MNRAS.379.1343S, 
2008ApJ...685...83E, 2008ApJ...684..933G, 2013ApJ...768..104F,
2013A&A...551A.143K, 2013MNRAS.432.1367L, 2011A&A...532A..57S, 
2014A&A...562A..87E, 2014A&A...567A.144C, 2014MNRAS.445.4073C, 2015ApJ...800...80L}.

The contemporary cosmological paradigm tells us that the formation and evolution 
of the cosmic web is governed 
by gravitational attraction of the dark matter (DM), and repulsion 
of the dark energy (DE). 
Superclusters, which are  the largest galaxy systems in the Universe,
can be used to test  cosmological models in many different ways.
Their sizes, richness, masses, and frequencies 
have to be reproduced by cosmological models \citep{2006A&A...459L...1E, 
2011MNRAS.417.2938S, 2012ApJ...759L...7P},
their evolution may lead to  virialised structures in the future
\citep{2009MNRAS.399...97A, 2011MNRAS.415..964L, 2015A&A...575L..14C}, 
a few systems among them may
already have started collapsing \citep{1998ApJ...492...45S, 2000AJ....120..523R,
2002MNRAS.337.1417G, 2014MNRAS.441.1601P, 2015A&A...577A.144T}.
Nearby superclusters can be studied using galaxy velocity field
\citep{2014Natur.513...71T}.
The spatial distribution of superclusters can be used as a test
for characteristic scales in the cosmic web 
\citep{1994MNRAS.269..301E, 1997Natur.385..139E, 
2012ApJ...749...81H, 2015MNRAS.448.1660R}.
Superclusters may embed large amounts of warm-hot intergalactic medium
(WHIM) \citep[][and references therein]{2001ApJ...555..558R, 2014arXiv1410.1311N}.
Superclusters and supervoids can leave imprints on the 
cosmic  microwave background as hot and cold spots 
\citep[][and references therein]{2008ApJ...683L..99G}.

Galaxy superclusters have complex inner structures which can be 
quantified with morphological descriptors such as Minkowski functionals. 
The morphology of superclusters can be used as a test for cosmological models
\citep{2002MNRAS.331.1020K, 2007A&A...462..397E, 2007A&A...476..697E}.
The morphology of superclusters has been studied, for example,  by
\citet{2011MNRAS.411.1716C} and \citet{2007A&A...476..697E}. 
\citet{2007A&A...476..697E, 2011A&A...532A...5E} found in the large 
morphological variety of superclusters two main types of superclusters,
filaments and spiders. In filaments high-density core(s) of superclusters
are connected by a small number of galaxy chains (filaments).  In spiders there is
a large number of galaxy chains between high-density cores in
superclusters. Poor spider-type superclusters are similar to our Local
Supercluster with one rich cluster and filaments of galaxies and poor groups of
galaxies around it \citep{2007A&A...476..697E}.  
\citet{2014A&A...562A..87E} showed that galaxy content of filament- and spider-type
superclusters differ: filament-type systems contain relatively more 
red, passive galaxies than spider-type systems.
A large morphological variety of superclusters and their galaxy content
has not yet reproduced by cosmological models 
\citep{2007A&A...476..697E, 2011ApJ...736...51E}.

Among the galaxy superclusters analysed in \citet{2011A&A...532A...5E}, 
the supercluster SCl~001 from \citet{2012A&A...539A..80L} list 
of superclusters (we call it the 
A2142 supercluster because of the brightest galaxy cluster
in the supercluster, Abell cluster A2142)  has an exceptional shape and density. 
The main body of the supercluster is a $50$~\Mpc\  straight filament
of galaxy groups and clusters, it is the only supercluster
in the study by \citet{2011A&A...532A...5E} with such a shape. In addition, in this
supercluster  the value of the luminosity density in the Sloan Digital Sky Survey (SDSS)
is the highest \citep{2012A&A...539A..80L}. Exceptional objects are 
always interesting; they may represent a challenge for cosmological
models \citep{2007A&A...476..697E, 2011MNRAS.417.2938S, 2012ApJ...759L...7P}.  

The goal of the present study is to analyse the structure of the 
A2142 supercluster as determined by galaxy groups, clusters, filaments and single
galaxies. We analyse the global density distribution in the supercluster,
divide the supercluster into a high-density core and lower-density outskirts
regions, and analyse the properties of these regions. 
We apply 1D and 3D tests as estimators of the dynamical
state of groups and clusters in regions of different global density.
We use a spherical collapse model to find whether this supercluster
may already be collapsing. 
In a forthcoming paper we shall analyse the galaxy populations in 
the A2142 supercluster. This is the first detailed study of the A2142 supercluster.

At \url{http://www.aai.ee/~maret/SCl1structure.html}
we present an interactive 3D model showing the distribution of
galaxies in the regions of different global density in the A2142 supercluster.

We assume  the standard cosmological parameters: the Hubble parameter $H_0=100~ 
h$ km~s$^{-1}$ Mpc$^{-1}$, the matter density $\Omega_{\rm m} = 0.27$, and the 
dark energy density $\Omega_{\Lambda} = 0.73$ \citep{2011ApJS..192...18K}.

\section{Data} 
\label{sect:data} 

We use the MAIN sample of the 8th and 10th data release of the Sloan Digital Sky 
Survey \citep{2011ApJS..193...29A, 2014ApJS..211...17A} with the 
apparent Galactic extinction corrected $r$ magnitudes $r \leq 
17.77$, and the redshifts $0.009 \leq z \leq 0.200$. 
We corrected the redshifts of galaxies for the motion relative to the CMB and 
computed the comoving distances \citep{2002sgd..book.....M} of galaxies. 
For details of sample selection we refer to \citet{2012A&A...539A..80L},
\citet{2012A&A...540A.106T}, and  \citet{2014A&A...566A...1T}.

\subsection{Luminosity density field and superclusters}
\label{subsec:scl}

We calculate the galaxy luminosity density field to reconstruct the underlying 
matter distribution, and to determine superclusters 
(extended systems of galaxies) in the luminosity density field
at the smoothing length 8~\Mpc\ using $B_3$ spline kernel.
We created a set 
of density contours by choosing a density thresholds and defined connected 
volumes above a certain density threshold as superclusters. In order to choose 
the proper density level for determining individual superclusters, we analysed the 
properties of the density field superclusters at a series of density levels. 
As a result we used 
the density level $D8 = 5.0$
(in units of mean density, $\ell_{\mathrm{mean}}$ = 
1.65$\cdot10^{-2}$ $\frac{10^{10} h^{-2} L_\odot}{(\vmh)^3}$)
to determine individual superclusters.
At this density level superclusters in the richest 
chains of superclusters in the volume under study  still form separate systems; 
at lower density levels they join into huge percolating systems. At higher 
threshold density levels superclusters are smaller and 
there are fewer superclusters.
The calculation of the luminosity density field and determination of
superclusters is described in detail in   \citet{2012A&A...539A..80L}. 

The  morphology of superclusters up to a distance of $340$~\Mpc\ was determined  
in \citet{2012A&A...542A..36E} who  
classified superclusters as filaments and spiders on the basis of their morphological
information and visual appearance.
The details about supercluster
morphology are given in \citet{2007A&A...476..697E, 2011A&A...532A...5E}.

\begin{figure}[ht]
\centering
\resizebox{0.48\textwidth}{!}{\includegraphics[angle=0]{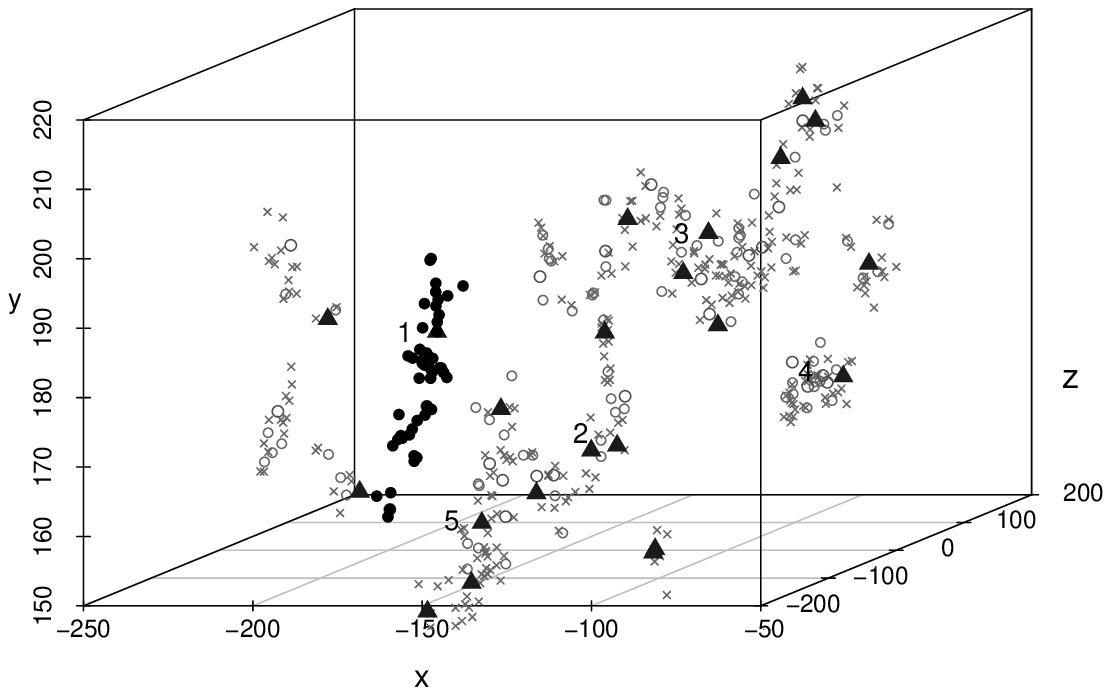}}
\caption{
Distribution of galaxy groups with at least four member galaxies
in cartesian coordinates (in $h^{-1}$~Mpc) in the neighbourhood of the A2142 supercluster. 
Black filled triangles show groups with 
at least 50 member galaxies, empty circles groups
with $10 \leq N_{\mathrm{gal}} < 50$, 
and crosses groups with less than ten galaxies. 
Black filled circles show galaxy groups in the A2142 supercluster with at least 
four member galaxies.
Numbers denote some rich superclusters from
\citet{2012A&A...539A..80L} list 
of superclusters: 1 marks the A2142 supercluster
(SCl~001), 
2 the Corona Borealis supercluster, 3  the richest supercluster 
in the Sloan Great Wall, 4  the Bootes supercluster 
\citep{2012A&A...539A..80L}, and 5  a rich supercluster separated
from the A2142 supercluster and the Corona Borealis supercluster by a void. 
The $x$, $y$, and $z$ coordinates are
defined as in 
\citet{2007ApJ...658..898P} and in \citet{2012A&A...539A..80L}:
$x = -d \sin\lambda$, $y = d \cos\lambda \cos \eta$, and
$z = d \cos\lambda \sin \eta$,
where $d$ is the comoving distance, and $\lambda$ and $\eta$ are the SDSS 
survey coordinates.
}
\label{fig:scl}
\end{figure}

\begin{figure}[ht]
\centering
\resizebox{0.44\textwidth}{!}{\includegraphics[angle=0]{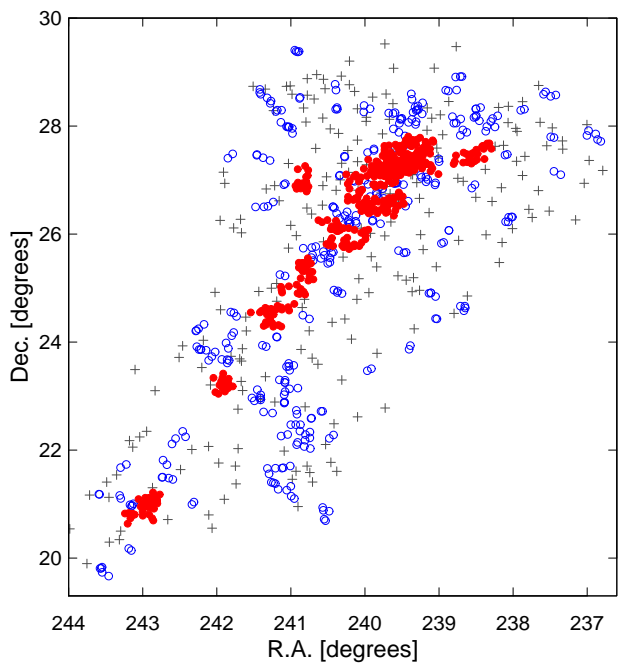}}
\caption{
Sky distribution of galaxies in the A2142 supercluster. 
Red filled circles show galaxies in groups and clusters with 
at least ten member galaxies, blue empty circles  galaxies in groups
with $2 \leq N_{\mathrm{gal}} \leq 9$, and grey crosses  single
galaxies. 
}
\label{fig:radec}
\end{figure}

In Table~\ref{tab:scldata} we present data about the A2142 supercluster,
classified as a filament-type in \citet{2011A&A...532A...5E}.
The maximum global density in A2142 supercluster is $D8 = 21.6$,
this is the highest in the whole SDSS volume 
\citep[in][the volume used to determine superclusters
was $V = 0.132~(\vgh)^{3}$]{2012A&A...539A..80L}. 
Figure~\ref{fig:scl} shows
the location of the A2142 supercluster among neighbouring relatively rich
superclusters. The A2142 supercluster is surrounded by voids. The nearest
rich supercluster to the A2142 supercluster is 
the Corona Borealis supercluster  at the distance of about $40$~\Mpc.
The distribution of rich superclusters
was described in more detail in \citet{2011A&A...532A...5E}.
In Fig.~\ref{fig:radec} we plot
the sky distribution of galaxies in the A2142 supercluster. 
Figure~\ref{fig:radec} shows that galaxy groups and clusters 
in this supercluster lie along an almost straight line
on the plane of the sky. Such a straight
filament as the main body of the supercluster is unique among superclusters
analysed in \citet{2011A&A...532A...5E}.

\begin{table*}[ht]
\caption{General properties of A2142 supercluster.}
\label{tab:scldata}  
\begin{tabular}{rrrrrrrr} 
\hline\hline 
\multicolumn{1}{c}{(1)}&(2)&(3)&(4)& (5)&(6)&(7)&(8)\\      
\hline 
\multicolumn{1}{c}{ID}& $N_{\mbox{gal}}$ &$N_{\mbox{gr}}$ &$N_{\mbox{10}}$& $d_{\mbox{peak}}$  
 & $L_{\mbox{tot}}$ & $\mathrm{Diam}$ & $D8_{\mathrm{max}}$ \\
& &  & & [$h^{-1}$ Mpc] & [$10^{10}h^{-2} L_{\sun}$] & [$h^{-1}$ Mpc] & \\
\hline
 239+027+009 & 1038  & 108 & 14 & 264.5  &  1591.5&  50.3 & 21.6 \\ 
\hline
\end{tabular}\\
\tablefoot{                                                                                 
Columns in the Table are as follows:
(1): the supercluster ID AAA+BBB+ZZZ, where AAA is R.A., +/-BBB is Dec., and ZZZ is 100$z$;
(2): the number of galaxies in the supercluster, $N_{\mbox{gal}}$;
(3): the number of groups in the supercluster, $N_{\mbox{gr}}$;
(4): the number of groups with at least 10 member galaxies in the supercluster, $N_{\mbox{10}}$;
(5): the distance of the density maximum, $d_{\mbox{peak}}$;
(6): the total weighted luminosity of galaxies in the supercluster, $L_{\mbox{tot}}$;
(7): the supercluster diameter (the maximum distance between galaxies in
the supercluster), $\mathrm{Diam}$;
(8): the maximal value of the luminosity-density field calculated with
the $8$~\Mpc\ smoothing kernel, $D8_{\mathrm{max}}$, in units of the mean density as described in the text.
}
\end{table*}

\begin{figure}[ht]
\centering
\resizebox{0.44\textwidth}{!}{\includegraphics[angle=0]{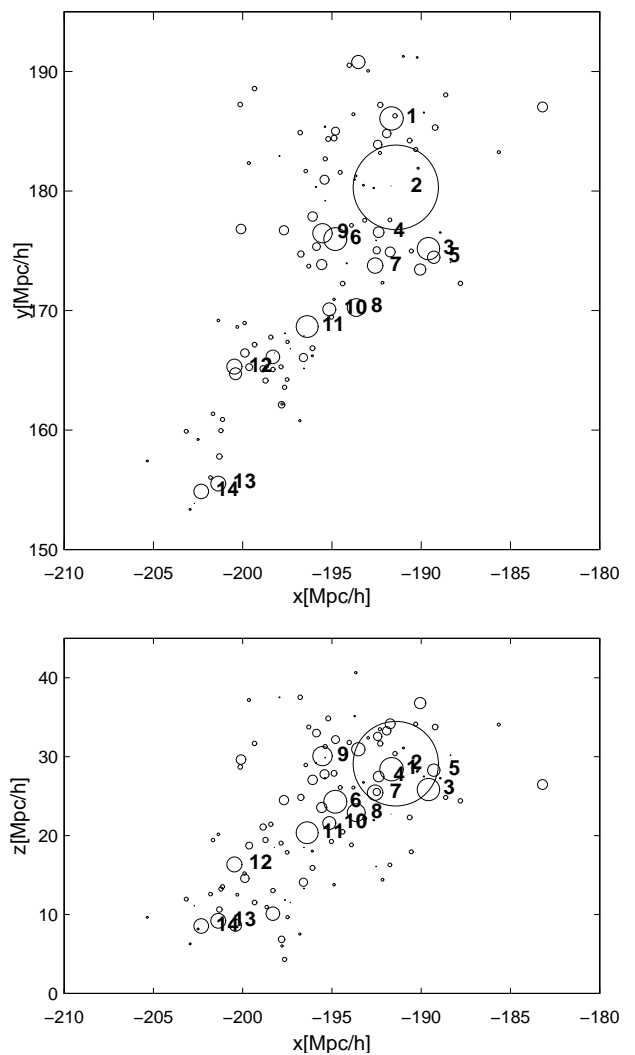}}
\caption{
Distribution of galaxy groups  in the  A2142 supercluster
in cartesian coordinates. 
Circle sizes are proportional to each group's size in the sky,
numbers are order numbers of groups with at least ten member galaxies
from Table~\ref{tab:gr10}.
}
\label{fig:radecgr}
\end{figure}

\subsection{Galaxy groups}
\label{subsec:gr}

We selected galaxy groups belonging to the 
A2142 supercluster from the group catalogue 
based on the 10th data release of the Sloan Digital Sky 
Survey by \citet{2014A&A...566A...1T}.
In this catalogue galaxy groups were  determined using 
the friends-of-friends cluster analysis 
method introduced in cosmology by \citet{1982Natur.300..407Z} and 
\citet{1982ApJ...257..423H}. A galaxy belongs to a 
group of galaxies if this galaxy has at least one group member galaxy closer 
than a linking length. In a flux-limited sample the density of galaxies slowly 
decreases with distance. To take this selection effect into account properly 
when constructing a group catalogue from a flux-limited sample, the 
linking length was rescaled with distance, calibrating the scaling relation by observed 
groups. As a result, the 
maximum sizes in the sky projection and the velocity dispersions of our groups 
are similar at all distances. The details about the data 
reduction, the group finding procedure, and the description of the group catalogue can be found in 
\citet{2014A&A...566A...1T}.

The A2142 supercluster embeds 14 galaxy groups with at least ten member
galaxies. We list them in Table~\ref{tab:gr10}. The richest group, Gr~3070, corresponds
to the Abell cluster A2142. 
Figure~\ref{fig:radecgr}  shows
the distribution of galaxy groups in cartesian coordinates. 
Galaxy groups in the A2142 supercluster cover only a narrow distance interval 
($\approx 255-275$~\Mpc, Table~\ref{tab:gr10}); 
therefore, we use group data from flux-limited group catalogue.

\begin{table*}[ht]
\caption{Data on groups in the A2142 supercluster with at least ten member galaxies.}
\begin{tabular}{rrrrrrrrrrrrrr} 
\hline\hline  
(1)&(2)&(3)&(4)&(5)& (6)&(7)&(8)&(9)&(10)&(11)&(12) \\      
\hline 
No. & ID&$N_{\mathrm{gal}}$& $\mathrm{R.A.}$ & $\mathrm{Dec.}$ 
&$\mathrm{Dist.}$  & $\sigma_{\mathrm{v}}$ & $r_{\mathrm{max}}$ & $L_{\mathrm{tot}}$  
& $M_{\mathrm{dyn}}$ & $M/L$ & $D8$  \\
&&&[deg]&[deg]&[$h^{-1}$ Mpc]&$[{\mathrm{km~s^{-1}}}]$&[$h^{-1}$ Mpc]& [$10^{10} h^{-2} L_{\sun}$] 
 & [$10^{12}h^{-1}M_\odot$] & [$hM_\odot$/$L_\odot$] &  \\
\hline
 1 & 10570 &  27 & 238.53 & 27.47 &  268.57 &  247.76  & 1.26 & 51.3 &   59.5& 116.2 & 13.65 \\
 2 &  3070 & 212 & 239.52 & 27.32 &  264.64 &  769.02  & 2.39 &382.0 &  906.5& 237.3 & 20.70 \\
 3 &  4952 &  54 & 239.78 & 26.56 &  260.11 &  447.24  & 1.30 &111.0 &  214.4& 193.1 & 17.13 \\
 4 & 32074 &  11 & 240.11 & 26.71 &  262.42 &  337.05  & 0.61 & 15.7 &   59.5& 378.2 & 19.89 \\
 5 & 35107 &  10 & 240.13 & 27.01 &  258.69 &  221.02  & 0.77 & 13.9 &   30.3& 218.4 & 14.20 \\
 6 & 14960 &  27 & 240.20 & 25.87 &  263.58 &  382.58  & 1.30 & 46.6 &  147.5& 316.6 & 15.32 \\
 7 & 17779 &  20 & 240.38 & 26.16 &  261.06 &  371.74  & 1.15 & 35.1 &  104.1& 296.9 & 16.39 \\
 8 &  6885 &  32 & 240.75 & 25.40 &  258.85 &  428.12  & 1.22 & 70.5 &  164.1& 232.8 & 11.38 \\
 9 & 21183 &  21 & 240.83 & 26.95 &  265.02 &  280.10  & 1.29 & 36.5 &   61.2& 167.6 & 15.20 \\
10 & 20324 &  11 & 240.95 & 24.95 &  259.72 &  335.77  & 0.73 & 17.4 &   77.5& 446.2 & 10.42 \\
11 & 10818 &  28 & 241.23 & 24.52 &  259.56 &  328.02  & 1.29 & 51.3 &  105.1& 205.1 &  9.31 \\
12 & 14283 &  19 & 241.90 & 23.21 &  260.28 &  323.03  & 0.86 & 34.4 &   65.1& 189.0 &  7.27 \\
13 & 10224 &  32 & 242.91 & 21.02 &  254.55 &  633.40  & 0.86 & 60.4 &  232.5& 385.0 &  6.01 \\
14 & 26895 &  12 & 243.08 & 20.77 &  254.80 &  257.42  & 0.92 & 16.5 &   49.1& 297.9 &  5.91 \\
\hline
\label{tab:gr10}  
\end{tabular}\\
\tablefoot{                                                                                 
Columns are as follows:
(1): Order number of the group;
(2): ID of the group from \citet{2014A&A...566A...1T} (Gr~3070
correspond to the Abell cluster A2142);
(3): the number of galaxies in the group, $N_{\mathrm{gal}}$;
(4)--(5): group center right ascension and declination;
(6): group center comoving distance;
(7): rms velocity of galaxies in the group;
(8): group maximum radius;
(9): group total luminosity;
(10): dynamical mass of the group assuming the NFW density profile, $M_{\mathrm{dyn}}$;
(11): group mass-to-light ratio $M/L$;
(12): the value of the luminosity-density field at the location of the group calculated with
the $8$~\Mpc\ smoothing length, $D8$, in units of the mean density as described in the text.
}
\end{table*}

\subsection{Galaxy filaments}
\label{subsec:fil}

\begin{figure}[ht]
\centering
\resizebox{0.44\textwidth}{!}{\includegraphics[angle=0]{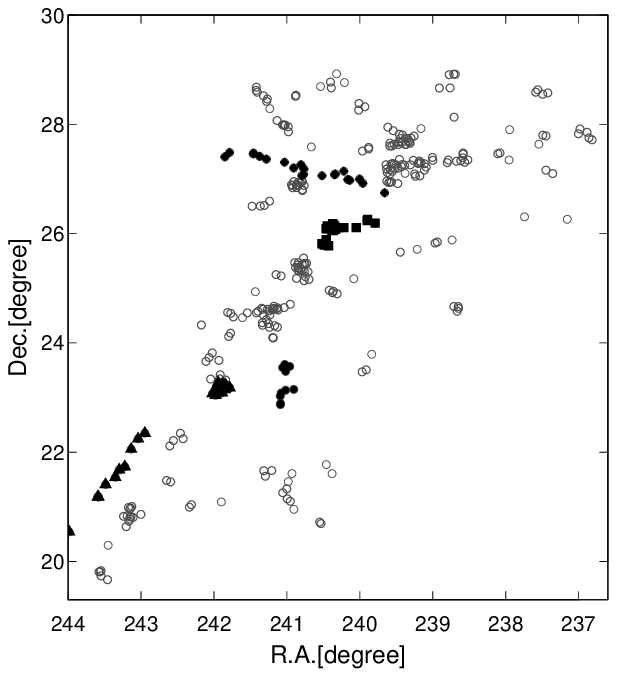}}
\caption{
Distribution of galaxies in filaments in the A2142 supercluster
in the sky (empty circles). 
Black filled symbols show members of filaments F~03116 (diamonds), 
F~03894 (squares), F~00580 (circles), and F~04462
(triangles).
}
\label{fig:radecfil}
\end{figure}

We use the catalogue of galaxy filaments from the flux-limited
filament catalogue by \citet{2014MNRAS.438.3465T}
that was built by applying the Bisous process
to the distribution of galaxies as outlined in \citet{2014MNRAS.438.3465T}.
From this catalogue we extracted galaxy filaments
which had at least one member galaxy from the A2142 supercluster
with distance $d \leq 0.8$~\Mpc\ from the filament axis, as defined for 
filament members in \citet{2014MNRAS.438.3465T}. From this
initial sample we included in our sample of A2142 filaments those
with member galaxies with global density $D8 \geq 4.8$.
With this limit we excluded those filaments that were outgoing from
the supercluster to the surrounding voids.  
Our final filament list contains 31 filaments, four which are longer than $10$~\Mpc:
F00580 with the length of $11.1$~\Mpc, F03116 with the length of $17.2$~\Mpc,
F03894 with the length of $18.1$~\Mpc, and
F04462 with the length of $19.1$~\Mpc.
The shortest galaxy filaments in the A2142 supercluster are $2$~\Mpc\
long; the median length of filaments shorter than $10$~\Mpc\
is $5.6$~\Mpc. In total, approximately one-third of all galaxies in the A2142 supercluster
are filament members. The distribution of galaxies in filaments
in the sky is plotted in Fig.~\ref{fig:radecfil}. 
Figure~\ref{fig:radecfil} shows filaments radially going out of the central
part of the supercluster. This is more clearly seen from
the 3D model at our web pages. We plan to study this feature and 
other properties of galaxy filaments in the A2142 supercluster in more detail in a separate paper.

Catalogues of galaxy superclusters, groups and filaments used in this study are
available from the database of cosmology-related catalogues at \url{http://cosmodb.to.ee/}.

\section{Methods}
\label{sect:methods} 

\subsection{Multidimensional normal mixture modelling}
\label{sect:mclust}

To search for possible components (substructure) 
in clusters, we employ multidimensional normal 
mixture modelling based on the analysis of a finite mixture of distributions, in 
which each mixture component is taken to correspond to a different component
or subgroup in sky coordinates and velocity of galaxies in groups and clusters. 
The most common component distribution considered in 
model-based clustering is a multivariate Gaussian (or normal) distribution.  To 
model the collection of components, we apply the {\it 
Mclust} package for classification and clustering
\citep{fraley2006} from {\it R}, an open-source free statistical environment 
developed under the GNU GPL \citep[][\texttt{http://www.r-project.org}]{ig96}. 
This package searches for an 
optimal model for the clustering of the data among models with varying shape, 
orientation, and volume, finds the optimal number of components and the 
corresponding classification (the membership of each component). For every galaxy,
{\it Mclust} calculates  the probability of it belonging to any of 
the components. The uncertainty of classification is defined as one minus the 
highest probability that a galaxy belongs to a
component. The mean uncertainty 
for the full sample is used  as a statistical estimate of the reliability of the 
results.

In \citet{2012A&A...540A.123E}
we tested how the possible errors in the line-of-sight positions of galaxies 
affect the results  of {\it Mclust}, randomly shifting the peculiar velocities 
of galaxies 1000 times and searching  each time for the components with {\it 
Mclust}. The random shifts were chosen from a Gaussian
distribution with the dispersion equal to the velocity dispersion of galaxies 
in a cluster.  The number of the components found by {\it Mclust} remained unchanged, 
demonstrating that the results of {\it Mclust} are not sensitive to such errors.

\subsection{One-dimensional tests}

One indicator of multimodality in clusters is the deviation of the distribution 
of the galaxy velocities in clusters  from a Gaussian. 
An early use of the Gaussianity test was by \citet{1952PTarO..32..231E, 1954PTarO..32..371E}
where it was applied to test the homogeneity of stellar populations
in our Galaxy.  
We tested the hypothesis of the Gaussian 
distribution of the peculiar velocities of galaxies in clusters with the 
Shapiro-Wilk normality test \citep{shapiro65}, which is 
considered the best for small samples, and with the Anderson-Darling
test, which is very reliable according to \citet{2009ApJ...702.1199H}.
To test for the 
asymmetry of the distributions of galaxy velocities 
we calculated the kurtosis of the peculiar velocity distribution using
the Anscombe-Glynn test \citep{anscombe83} 
from the $R$ package {\it moments} by L. Komsta and F. Novomestky.
Kinematic models show that 
the shape of the velocity distribution of galaxies in groups
is defined by the ratio of different 
types of galaxy orbits \citep{1987ApJ...313..121M}. The tests described here 
serve mostly to check if the galaxy velocity distribution is unimodal and symmetrical. 

In virialised clusters galaxies follow the potential well of the cluster. We 
would expect that the main galaxies in clusters lie at the centres of 
groups (group haloes) and have small peculiar velocities 
\citep{1975ApJ...202L.113O,1984ApJ...276...26M,1992ApJ...386..420M}. 
Therefore we calculated the peculiar velocities of the main galaxies 
$|v_{\mathrm{pec}}|$ and the normalised peculiar velocities of the main 
galaxies, $v_{\mathrm{pec,r}} = |v_{\mathrm{pec}}|/{\sigma}_{v}$,
where ${\sigma}_{v}$ is group rms velocity. 
In the group catalogue the main galaxy of a group is defined
as the most luminous galaxy in the $r$-band. We also use this definition 
in the present paper. 
\citet{2012A&A...540A.123E} showed for very rich groups with $N_\mathrm{gal} \geq 50$
that in groups with the peculiar velocities of the main galaxies 
$|v_{\mathrm{pec}}| < 250~{\mathrm{km~s^{-1}}}$ ($|v_{\mathrm{pec,r}}| < 0.5$)
main galaxies are located near the centre of the group. 
Higher peculiar velocities of the 
main galaxies suggest that the main galaxy is located far from the centre, a 
sign of a possible dynamically young group.

Recently \citet{2013MNRAS.434..784R}
used several methods to analyse the structure and dynamical state of 
galaxy groups and showed that these methods 
give quite reliable results for groups with more than ten member galaxies. 
To be on the safe side, we applied 
3D normal mixture modelling to groups with $N_{\mathrm{gal}} \geq 20$,
and used groups with $N_{\mathrm{gal}} \geq 10$ in 1D tests.

\section{Results}
\label{sect:results} 

\subsection{Luminosity density distribution in the A2142 supercluster}
\label{sect:d8} 

\begin{table*}[ht]
\caption{Galaxies, galaxy groups and filaments in the core and in
outskirts of A2142 supercluster.}
\begin{tabular}{rrrrrrrrrrr} 
\hline\hline  
(1)&(2)&(3)&(4)&(5)& (6)&(7)&(8)&(9)&(10)&(11)\\      
\hline 
 $D8$ & $N_{\mathrm{gal}}$& $f_{\mathrm{1}}$ &$N_{\mathrm{gr10}}$ & $N_{\mathrm{gr2-9}}$ & 
$N_{\mathrm{fil}}$ &  $M_{\mathrm{dyn}}$&  $M_{\mathrm{est}}$ & $L_{\mbox{gr}}$ & $M/L$& $M_{\mathrm{est}}/L$ \\
   & &&&&  & [$10^{15}h^{-1}M_\odot$]& [$10^{15}h^{-1}M_\odot$] & [$10^{12}h^{-2} L_{\sun}$] 
   & [h$M_\odot$/$L_\odot$] & [h$M_\odot$/$L_\odot$] \\
\hline
Full scl     &1038 & 0.21 &14 & 93 &  31 & 2.90 & 4.33 & 15.1  & 192 & 287\\  
 $ \geq 17$  & 300 & 0.03 & 3 &  8 &   1 & 1.21 & 1.38 &  5.3  & 226 & 260\\
 $ 13-17$    & 167 & 0.10 & 5 &  8 &   2 & 0.48 & 0.62 &  2.5  & 198 & 248\\
 $ 8-13$     & 197 & 0.29 & 3 & 24 &   5 & 0.54 & 0.90 &  2.9  & 186 & 310\\
 $ \leq 8$   & 374 & 0.35 & 3 & 53 &  11 & 0.67 & 1.41 &  4.3  & 156 & 398\\
\hline
\label{tab:D8prop}  
\end{tabular}\\
\tablefoot{
Columns are as follows:
(1): Global density $D8$;
(2): the number of galaxies in a region;
(3): the fraction of single galaxies among all galaxies in a region;
(4): the number of groups with at least ten member galaxies; 
(5): the number of groups with number of member galaxies $2 \leq N_{\mathrm{gal}} \leq 9$;
(6): the number of galaxy filaments fully embedded in a given region; 
(7): the total dynamical mass of groups (in case of groups with 2 and 3 member galaxies
we use median mass);
(8): the total estimated mass of a region (including faint groups and intracluster
    gas, see text);
(9): total luminosity of groups;
(10): mean mass-to-light ratio of groups in the region;
(11): mean estimated mass-to-light ratio of the region (see text for details).
}
\end{table*}

\begin{figure}[ht]
\centering
\resizebox{0.445\textwidth}{!}{\includegraphics[angle=0]{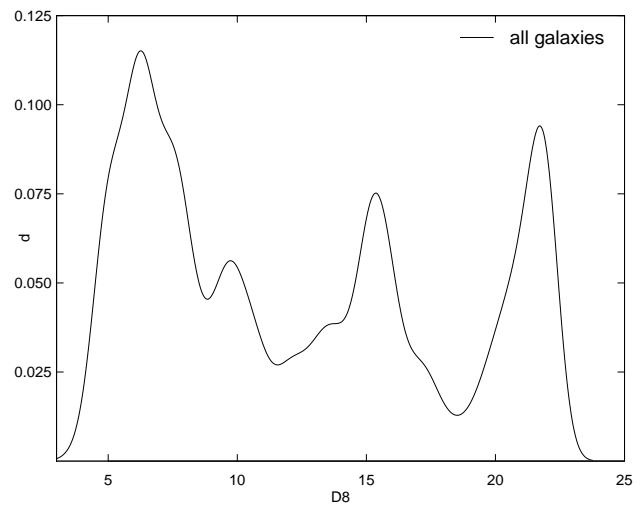}}
\caption{
Probability density distribution of global densities $D8$ around galaxies in A2142 supercluster. 
}
\label{fig:D8}
\end{figure}

\begin{figure}[ht]
\centering
\resizebox{0.44\textwidth}{!}{\includegraphics[angle=0]{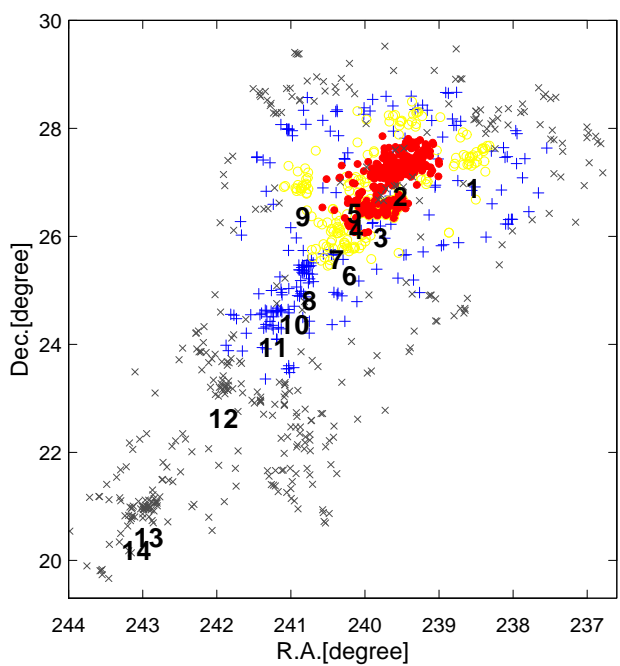}}
\caption{
Distribution of galaxies in different global 
density regions in the A2142 supercluster in the sky plane. 
Red filled circles denote galaxies in region of global
density $D8 \geq 17$, yellow empty circles correspond to galaxies with global density
$13 \leq D8 < 17$, blue crosses correspond to galaxies with global density
$8 \leq D8 < 13$, and grey Xs denote galaxies with $D8 < 8$.
Numbers are order numbers of groups with at least ten member galaxies
from Table~\ref{tab:gr10}.
}
\label{fig:radecd13}
\end{figure}

Figure~\ref{fig:D8} shows the probability density distribution
of the global densities $D8$ around galaxies in the A2142 supercluster.
This distribution shows several maxima and minima that can be used to
divide the supercluster into several global density regions.
Table~\ref{tab:D8prop} summarises the properties of 
these global density regions in A2142 supercluster, their
galaxy, group, and filament content, masses, and mass-to-light ratios. 
In Fig.~\ref{fig:radecd13} we show the sky distribution of galaxies in different 
global density regions. At \url{http://www.aai.ee/~maret/SCl1structure.html}
we present an interactive 3D version of Fig.~\ref{fig:radecd13}. 

In the highest density core of A2142 supercluster global density is as high as $D8 \geq 17$, 
the radius of this 
region is $\approx 5-6$~\Mpc. Here lie rich clusters A2142
(Gr~3070 in Table~\ref{tab:gr10}) and Gr~4952 with 54 galaxies. 
The region with $13 \leq D8 < 17$ and radius approximately $8$~\Mpc\ 
embeds five rich groups and clusters (see Table~\ref{tab:gr10}).
These regions form the high-density core of the supercluster which morphologically
resembles a spider with many short outgoing filaments.

In the intermediate global density region where $8 \leq D8 < 13$ lie 
three rich groups, the richest group has 28 member galaxies. 
Three rich groups lie in the outskirts 
of the supercluster with global density  $D8 < 8$.
The global density limit $D8 = 8$ is approximately the same
as the density limit between high-density cores of rich superclusters
and their outskirts \citep{2007A&A...464..815E}. 
Figure~\ref{fig:radecd13} shows that rich groups and clusters in the A2142 supercluster
lie along a straight line in the sky plane, forming the $50$~\Mpc\ long 
main body of the supercluster. The high-density core of the supercluster
occupies approximately one-third of the size of the supercluster, and two-thirds is covered by a
lower density "tail", therefore the supercluster as a whole
is classified as being of filament morphology \citep{2011A&A...532A...5E}.
The number of poor groups with less than ten galaxies
is more than three times higher in the outskirts than in the core region; 
these groups are mostly very poor, with two or three member galaxies.

Table~\ref{tab:D8prop} shows that the highest density core of A2142 supercluster 
contains almost 30\% of all galaxies from A2142 supercluster. Almost all galaxies here belong 
to galaxy groups; only 3\% of galaxies are single. At lower global densities
the fraction of single galaxies increases, and in the outskirts of  supercluster
they form about one-third of galaxy populations. 

In Table~\ref{tab:D8prop} we give the number of galaxy filaments in each
global density region. Here we list only those filaments that are fully inside a given
global density region. In our list of filaments twelve cross several global density regions.
For example, the $17$~\Mpc\ filament F03116 starts in the highest density
core of the supercluster and extends to the outskirts of the supercluster.
Other filaments longer than $10$~\Mpc\ lie in the global 
density region with $D8 < 13$ out
of high-density core of the supercluster. 
Table~\ref{tab:D8prop} shows that the number of galaxy filaments
increases in the outskirts of the supercluster. 

\subsection{Masses and mass-to-light ratios}
\label{sect:masses} 

To calculate total masses and mass-to-light ratios of galaxy groups 
in the regions of different global density we used dynamical masses  
of galaxy groups calculated in \citet{2014A&A...566A...1T}.
To calculate masses of galaxy groups \citet{2014A&A...566A...1T} used virial theorem, 
assuming symmetry of galaxy velocity distribution 
and the Navarro-Frenk-White (NFW)
density profile for galaxy distribution in the plane of the sky. 
For a detailed description how the dynamical masses of groups were calculated
we refer to \citet{2014A&A...566A...1T}.
For groups with 
$N_{\mathrm{gal}} = 2$ and $3$ we use
the median values of group masses. 
We also calculated stellar-to-dynamical mass ratios for groups
in the core and in the outskirts of the superclusters.
Data about the stellar masses of galaxies 
were taken 
from the MPA-JHU spectroscopic catalogue \citep{2004ApJ...613..898T, 
2004MNRAS.351.1151B} in the SDSS CAS database. In this catalogue the different properties of 
galaxies are obtained by fitting SDSS photometry and spectra with
the stellar population synthesis models developed by \citet{2003MNRAS.344.1000B}.
The stellar masses of galaxies are estimated from the 
galaxy photometry \citep{2003MNRAS.341...33K}. 

Table~\ref{tab:D8prop} presents sums of dynamical masses of galaxy groups
and mass-to-light ratios in the high-density core and the outskirts of the 
A2142 supercluster. 
Two mass values are given. First, we give the sums
of the dynamical masses of galaxy groups in a region.
In the second case we add to this the estimated mass
of faint galaxy groups as follows. 
In each region some galaxies are single. They may be the brightest galaxies
of faint groups in which other member galaxies are too faint to be 
observed within SDSS survey magnitude limits \citep{2009A&A...495...37T}.
We estimated the masses of these faint groups using the number of
single galaxies in a region, and median masses of groups
with less then five member galaxies. In addition, 
the mass of intracluster gas is about 10\% of the total mass 
\citep{2014arXiv1412.7823C}, we added this to the estimated mass 
and mass-to-luminosity ratio of each global density
region ($M_{\mathrm{est}}$  and $M_{\mathrm{est}}/L$ in Table~\ref{tab:D8prop}).

Table~\ref{tab:D8prop} gives the values of masses for $h = 1.0$,
as used in the present paper. The total estimated mass of the 
supercluster is $M_{\mathrm{est}} = 6.2\times~10^{15}M_\odot$ for 
$h = 0.7$ \citep{2015arXiv150201589P}.

Table~\ref{tab:D8prop} shows that the
highest density regions ($D8 \geq 13$) with their rich clusters contain most of the mass 
and most of the light of the supercluster, if we calculate masses using group dynamical masses. 
In the outskirts of the supercluster the estimated mass is approximately twice as large
as the sum of dynamical mass of galaxy groups. The supercluster total estimated
mass is approximately 1.5 times larger than the sum of dynamical mass of galaxy groups and
clusters. 
\citet{2014A&A...567A.144C} showed that the bias factor of the 
mass of superclusters constructed using galaxy groups and clusters 
is approximately 1.83. This suggests that we may still underestimate
the total mass of the supercluster.

The ratio of stellar masses to dynamical masses $M_{\mathrm{*}}/M_{\mathrm{dyn}}$
of groups is $0.017$ in the high-density core of the supercluster,
and $0.022$ in the outskirts of supercluster.  
Several studies have shown that 
stellar mass-to-total mass ratio increases toward lower halo masses;    
this may be the reason why this ratio is larger in the supercluster outskirts where
poor groups dominate than in the core
region of the supercluster \citep{2010MNRAS.407..263A, 2014MNRAS.439.2505B, 2015ApJ...799L..17P}.

Mean dynamical mass-to-light ratios have the
highest values in the high-density core of the supercluster, and the
lowest values in the outskirts where poor groups with lower
values of $M_{\mathrm{dyn}}/L$ dominate. Estimated mass-to-light
ratios $M_{\mathrm{est}}/L$ have higher values in the supercluster outskirts
where the fraction of single galaxies is the highest, and the
difference between the dynamical masses and estimated masses is the highest.

In Table~\ref{tab:gr10} we give dynamical masses, 
total luminosities, and mass-to-light
ratios for  galaxy groups with at least ten member galaxies. 

\citet{2014MNRAS.439.2505B} showed that mass-to-light ratios for groups
$M/L \approx 400$. In our sample only four groups have $M/L > 300$
(Table~\ref{tab:gr10}). These are either rather poor groups
with eleven member galaxies or groups with substructure for
which dynamical masses are not reliable. 
Mean dynamical group mass-to-light ratios are lower in
the supercluster outskirts regions with lower global density
(Table~\ref{tab:D8prop}). Eight 
groups with at least ten member galaxies lie in the 
high-density core of the supercluster while the low density outskirts is mainly 
populated by poor groups. So the decrease
in $M/L$ at lower global densities probably reflects group richness distribution,
in agreement with \citet{2014MNRAS.439.2505B} who showed that 
mass-to-light ratios of groups have lower values for poor groups, and 
does not depend on large-scale environment of groups.

We also calculated stellar masses of groups with at least ten member galaxies
and stellar-to-dynamical mass ratios. The median value of the 
stellar-to-dynamical mass ratio for these groups is $0.013$, in agreement
with other studies, which have found that the fraction of stellar
mass in groups and clusters is about 1\% of the total mass of clusters
\citep[][ and references therein]{2014arXiv1412.7823C}. 

Recently, different methods to calculate galaxy group and cluster masses, and their
uncertainties were discussed by 
\citet{2012MNRAS.426.1829N, 2015AJ....149...54T, 2015AJ....149..171T}. 
\citet{2014MNRAS.441.1513O} and \citet{2015MNRAS.449.1897O}   
compared several mass estimation methods using
simulated mock galaxy catalogues. The compared mass estimation methods can be divided
into several classes: abundance matching, richness-based methods, virial theorem,
various phase-space based methods, caustic technique 
\citep[see ][for references and more details about various methods]
{2014MNRAS.441.1513O, 2015MNRAS.449.1897O}. According to the comparison results the
virial theorem based method used in \citet{2014A&A...566A...1T} 
produces quite reliable masses with reasonable scatter compared
with the true masses of the galaxy groups/clusters.

\subsection{Dynamical state of groups}
\label{sect:dyn} 

Table~\ref{tab:dyn10} presents the results of the 3D and the 1D tests
for groups with at least ten member galaxies.
The number of components determined with the 3D normal
mixture modelling is denoted as $N_{\mathrm{comp}}$.
For the 1D test we give the statistical significance of the deviations 
from Gaussianity in their galaxy distributions 
(the $p$-values: $p \leq 0.05$
means statistically highly significant deviations from Gaussianity).
The $p$-values for the Anderson-Darling and Shapiro-Wilk Gaussianity tests are denoted,
correspondingly, $p_{\mathrm{AD}}$ and $p_{\mathrm{SW}}$. 
The $p$-value for the kurtosis test is $p_{\mathrm{kurt}}$.
We also give in this table the values of the peculiar velocities and
normalised peculiar velocities, $v_{\mathrm{pec}}$ and $v_{\mathrm{pec,r}}$.
The uncertainty of the 3D normal mixture modelling was
less than $10^{-2}$ for all clusters, we do not show this in the table. 

\begin{table}
\caption{Results of the tests for groups with at least ten member galaxies.}
\begin{tabular}{rrrrrrrr} 
\hline\hline  
(1)&(2)&(3)&(4)&(5)& (6)&(7)&(8)\\      
\hline 
 No. & $N_{\mathrm{gal}}$ & $v_{\mathrm{pec}}$ & $v_{\mathrm{pec,r}}$ &
$N_{\mathrm{comp}}$ & $p_{\mathrm{AD}}$ & $p_{\mathrm{SW}}$ & $p_{\mathrm{kurt}}$  \\
\hline
\multicolumn{8}{c}{High-density core of the supercluster}\\
\hline
 1 &  27 &  236.4 & 0.95 &  2 &  0.330 & 0.312 & 0.835  \\
 2 & 212 &  334.4 & 0.43 &  3 &  0.178 & 0.170 & 0.687  \\
 3 &  54 &  148.6 & 0.33 &  3 &  0.002 & 0.012 & 0.806  \\
 4 &  11 &  284.1 & 0.84 &  1 &  0.137 & 0.194 & 0.063  \\
 5 &  10 &  387.4 & 1.75 &  1 &  0.470 & 0.453 & 0.275  \\
 6 &  27 &  607.5 & 1.59 &  2 &  0.171 & 0.143 & 0.038  \\
 7 &  20 &  565.8 & 1.52 &  1 &  0.502 & 0.525 & 0.819  \\
 9 &  21 &  292.6 & 1.04 &  4 &  0.117 & 0.083 & 0.581  \\
\hline
\multicolumn{8}{c}{Low-density outskirts of the supercluster}\\
\hline
 8 &  32 &  659.3 & 1.54 &  1 &  0.764 & 0.623 & 0.892  \\
10 &  11 &  394.5 & 1.17 &  1 &  0.229 & 0.150 & 0.074  \\
11 &  28 &   55.7 & 0.17 &  3 &  0.409 & 0.494 & 0.475  \\
12 &  19 &  358.6 & 1.11 &  1 &  0.874 & 0.894 & 0.897  \\
13 &  32 &  651.1 & 1.03 &  3 &  0.024 & 0.028 & 0.002  \\
14 &  12 &  389.5 & 1.51 &  1 &  0.855 & 0.649 & 0.344  \\
\hline
\label{tab:dyn10}  
\end{tabular}\\
\tablefoot{
Columns are as follows:
(1): Number of a group from Table~\ref{tab:gr10};
(2): the number of galaxies in a group; 
(3): peculiar velocity of the main galaxy ($\mathrm{km~s^{-1}}$);
(4): normalised peculiar velocity of the main galaxy;
     $v_{\mathrm{pec}}$/$\sigma_{\mathrm{v}}$;
(5): the number of components in 3D in a group, $N_{\mathrm{comp}}$;
(6): $p$-value for AD test;
(7): $p$-value of SW test;
(8): $p$-value for $kurtosis$ test.
}
\end{table}

Table~\ref{tab:dyn10} shows that 
among galaxy groups with $N_{\mathrm{gal}} \geq 20$ in the high-density core of the
supercluster
seven are multicomponent groups suggesting that these groups
are dynamically active. Only one group here is unimodal (no. 7). 
This group lies in the outer region of the high-density
core of the supercluster. 
In the outskirts of the supercluster there are both unimodal and multimodal
groups. 
 
According to the 1D tests only two groups have
$p_{\mathrm{AD}} < 0.05$ and  $p_{\mathrm{SW}} < 0.05$ showing statistically highly
significant deviations from Gaussian distribution. 
Group Gr4952 (no. 3 in Table~\ref{tab:dyn10}) is located in supercluster
high-density core and Gr10224 (no. 13) in the outskirts of supercluster.

The two most Gaussian groups with the highest $p$-values from the 1D tests 
in the A2142 supercluster are groups Gr14283 (no. 12), and Gr26895 (no. 14) 
located out of supercluster high-density core with global density
$D8 = 7.27$ and $D8 = 5.91$ (Table~\ref{tab:gr10}).   

Only for one group in the high-density core of the supercluster
(no. 3 with 54 member galaxies in Table~\ref{tab:dyn10})
and for one group in the outskirts (no. 11 with 28 member
galaxies)
both the peculiar velocity and the normalised peculiar velocity
of its main galaxy 
have lower values than the limits 
$|v_{\mathrm{pec}}| < 250 {\mathrm{km~s^{-1}}}$ and
$|v_{\mathrm{pec,r}}| < 0.5$, which suggests that the 
main galaxy is located close
to the  centre of the group \citep{2012A&A...540A.123E}.
In all other groups the main galaxy has large peculiar and/or
normalised peculiar velocity, a possible sign that the groups
are still dynamically active. Even group no. 3 with three components
is not virialised yet; here the main galaxy lies in the central component.

\begin{figure}[ht]
\centering
\resizebox{0.44\textwidth}{!}{\includegraphics[angle=0]{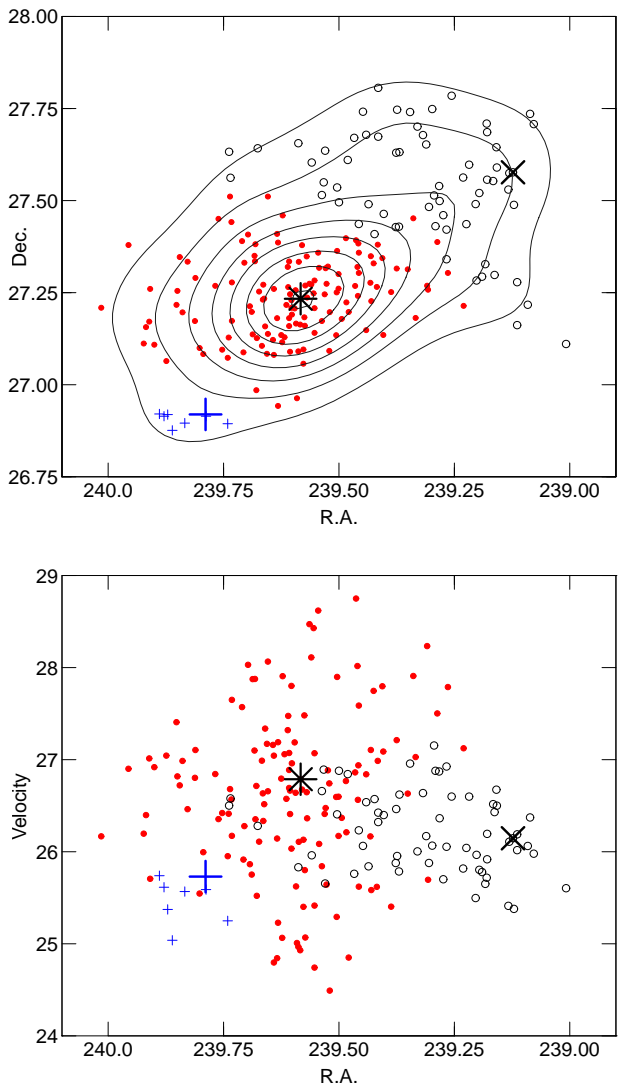}}\\
\caption{
Multimodal group Gr3070 (Abell cluster A2142, no. 2 in Table~\ref{tab:gr10}). 
Upper panel: sky distribution of galaxies in the group,
lower panel: R.A. vs. velocity (in $10^{3}~\mathrm{km~s^{-1}}$) plot; 
symbols show different galaxies belonging to different 
components as found by {\it Mclust}. 
The star marks the location of the main galaxy in the first component, 
the X shows the location of the brightest galaxy in the second component,
and the cross shows the location of the brightest galaxy in the third component.
Contours show the surface density distribution of galaxies in the whole cluster.
}
\label{fig:A2142}
\end{figure}

\begin{table}[ht]
\caption{Data for the Abell cluster A2142 (Gr3070) components.}
\begin{tabular}{rrrrrrr} 
\hline\hline  
(1)&(2)&(3)&(4)&(5)& (6)&(7)\\      
\hline 
 $ID$ & $N_{\mathrm{gal}}$&  $\mathrm{R.A.}$ & $\mathrm{Dec.}$  & $\mathrm{Dist.}$ 
 & $\sigma_{\mathrm{v}}$ &  $M_{\mathrm{dyn}}$ \\
 && [deg] & [deg] & [$Mpc~h^{-1}]$ &$[\mathrm{km~s^{-1}}]$ &\\
\hline
$A2142$ & 212 & 239.5 & 27.3 & 264.6 & 769.0 &  906 \\
 $C1$   & 135 & 239.6 & 27.2 & 266.3 & 851.9 &  707 \\
 $C2$   &  64 & 239.3 & 27.5 & 262.3 & 406.0 &  196 \\
 $C3$   &   8 & 239.8 & 26.9 & 254.9 & 235.7 &   15 \\
 \hline
\label{tab:a2142}  
\end{tabular}\\
\tablefoot{                                                                                 
Columns are as follows:
(1): Component ID;
(2): the number of galaxies in a component;
(3--5) R.A., Dec., and distance of a component centre; 
(6): rms velocity of galaxies in a component;
(7): the dynamical mass of a component, in $10^{12}h^{-1}M_\odot$. 
}
\end{table}

\subsection{Abell cluster A2142}
\label{sect:A2142}

In Fig.~\ref{fig:A2142} we show the distribution of galaxies
from three components detected with {\it Mclust}
in the Abell cluster A2142 (the richest group in the A2142 supercluster, Gr3070). 
We show also the overall density distribution of galaxies in the cluster.
We note that the orientation of density contour ellipses coincides 
with the orientation of the supercluster axis as determined by groups 
of galaxies (see, e.g. Fig.~\ref{fig:radecd13} and the other figures
showing sky distribution of galaxies and groups
in the A2142 supercluster).

The centre coordinates, velocity dispersions, and dynamical masses
of the components in this group are given in Table~\ref{tab:a2142}.
Five galaxies do not belong to any component. 
Table~\ref{tab:dyn10} shows that
1D tests did not detect  statistically significant deviations from
Gaussianity, and the reason can be seen in Fig.~\ref{fig:A2142}, where
the distribution of galaxy velocities in the components of this group
partially overlap and cannot be clearly distinguished. 

The main galaxy of the group is located in
the centre of the main component.
In the second and third components the brightest galaxy lies near the edge of the 
component (Fig.~\ref{fig:A2142}). 
 
A2142 is a well-known cold front X-ray cluster with a radio halo
that shows signs of merging and sloshing 
\citep{2000ApJ...541..542M, 2001cghr.confE..33F, 2011ApJ...741..122O, 
2013A&A...556A..44R, 2013ApJ...779..189F, 2014A&A...566A..68M}.
\citet{2013A&A...556A..44R} describe A2142 as a unique cluster
with giant radio halo where the sloshing occurs at much
larger scales compared to simulations and to most clusters with
sloshing cold fronts, up to about $1$~\Mpc. 
X-ray studies show a complex substructure in this cluster 
\citep{2008PASJ...60..345O, 2009ApJ...694.1643U}. 
\citet{2013A&A...556A..44R} suggest that the cluster A2142 
shows indication for a merger mainly along the line
of sight, which could trigger the radio emission without significantly
disturbing the X-ray morphology. 

Comparison with the X-ray and radio studies shows that  the orientation
of optical components of the cluster A2142 (Fig.~\ref{fig:A2142})
follows the orientation  of the radio halo and X-ray substructure,
as seen, for example, from Fig.~14 in \citet{2011ApJ...741..122O}, Fig.~7 in 
\citet{2013A&A...556A..44R} and Fig.~11 in \citet{2013ApJ...779..189F}.
In addition, the orientation of the cluster A2142 follows the orientation of
the whole supercluster. The similarity of 
orientations is related to the
very different scales (approximately $1-2$~\Mpc\ of
X-ray and radio haloes, $2.4$~\Mpc\ (A2142 maximum radius)   of the cluster
itself,  and about $50~$~\Mpc\ of the full supercluster).

\citet{2014A&A...566A..68M} found that mass profiles of  A2142
as determined by red and blue galaxies are different. This may be due to the different
galaxy content of A2142 components. We analyse this in a forthcoming paper.

Several X-ray and other studies studies have found that the mass of A2142 
is approximately 
$M = 0.9\times~10^{15}h^{-1}M_\odot$
\citep{2008PASJ...60..345O, 2009ApJ...694.1643U, 2010MNRAS.408.2442W,
2014A&A...566A..68M, 2015ApJ...800..122S}. \citet{2014A&A...566A..68M}
compared several mass estimation methods and concluded that the mass
of the cluster is $M_{200} = 0.875\times~10^{15}h^{-1}M_\odot$ (if recalculated to 
the Hubble parameter $H_0=100~h$~km~s$^{-1}$~Mpc$^{-1}$, as used in our paper); 
this value is 0.965 times lower than the dynamical mass of A2142 as calculated in
\citet{2014A&A...566A...1T}. This shows that the masses in \citet{2014A&A...566A...1T}
are quite precise.

Recently, \citet{2014A&A...570A.119E} reported the discovery
of an infalling group of galaxies in A2142 at  
$R.A. = 239.75$~degrees and $Dec = 27.41$~degrees. 
Galaxies in this group probably belong to the main component
of this cluster. Another possible 
merging group near the group Gr10570 was detected by \citet{2011ApJ...727L..38K}.

We also calculated stellar masses of galaxies in components $M_{\mathrm{*}}$, and the ratios
of stellar-to-dynamical masses, $f_{\mathrm{*}} = M_{\mathrm{*}}/M_{\mathrm{dyn}}$. 
The ratios $f_{\mathrm{*}} = 0.013, 0.017$, and $0.040$ in the first, second, and
third component, respectively. This is in agreement with the studies
of galaxy groups which have found that 
stellar mass-to-total mass ratio decreases toward higher group masses 
(see references in the previous section).

\section{Discussion and conclusions}
\label{sect:discussion}

\subsection{A2142 supercluster: a collapsing system?}
\label{sect:collapse} 

We applied the spherical collapse model to study the dynamical state 
of the A2142 supercluster. This model has been used to study the superclusters 
in several papers \citep{1998ApJ...492...45S, 2002MNRAS.337.1417G, 
2000AJ....120..523R, 2006A&A...447..133P, 2009MNRAS.399...97A, 2014MNRAS.441.1601P}. 
The spherical collapse model describes the evolution of a spherically symmetric 
perturbation in an expanding universe. The first essential moment in this evolution 
is called turnaround, the moment when the sphere stops expanding together with the universe
and the collapse begins. The turnaround 
marks the epoch when the perturbation decouples entirely from the Hubble flow 
of the homogeneous background.  
\citet{2002MNRAS.337.1417G} studied the dynamical state of 
superclusters in different $\Lambda$CDM models and showed that at the 
supercluster scale the turnaround is rare. Only a small fraction of superclusters 
or their high-density cores have already reached the turnaround radius and 
have started to collapse. In our nearby universe it may have happened, for example, 
in the Corona Borealis supercluster \citep{1998ApJ...492...45S, 2014MNRAS.441.1601P}, 
in the Shapley supercluster \citep{2000AJ....120..523R, 2006A&A...447..133P},
and in the A2199 supercluster \citep{2002AJ....124.1266R}
\citep[classified as the member of the Hercules supercluster in][]{1975ATsir.895....2E, 2001AJ....122.2222E}.

The density perturbation in the volume $V$ can be calculated as
\begin{equation}
\Delta\rho = \rho/\rho_{\mathrm{m}},
\end{equation} 
where  $\rho = M/V$ is the matter density in this volume and 
$\rho_{\mathrm{m}} = \Omega_{\mathrm{m}}\rho_{\mathrm{crit}} =
 3 \Omega_{\mathrm{m}} H_0^2 / 8\pi G$ 
is the mean matter density in the local universe. 
For a spherical volume $V = 4\pi/3 R^3$, we find that
\begin{equation}
\Delta\rho = 0.86\times 10^{-12}\Omega_{\mathrm{m}}^{-1}~(\frac{M}{h^{-1}M_\odot})~(\frac{R}{h^{-1}~{\mathrm{Mpc}}})^{-3}.
\label{eq:sph}
\end{equation} 
Assuming $\Omega_{\mathrm{m}} = 0.27$, as adopted in the present paper, 
the density fluctuation can be written as
\begin{equation}
\Delta\rho = 3.19\times 10^{-12}~(\frac{M}{h^{-1}M_\odot})~(\frac{R}{h^{-1}~{\mathrm{Mpc}}})^{-3}.
\label{eq:fluc}
\end{equation} 

According to the spherical collapse model, 
if $\Delta\rho > \Delta\rho_{\mathrm{T}}$ then
the perturbed region ceases to expand and begins to collapse. 
The values of turnaround parameters for various cosmological models
were given in \citet{2015A&A...575L..14C}. 
For $\Omega_{\mathrm{m}} = 0.27$ and $\Omega_{\mathrm{\Lambda}} = 0.73$, 
the density perturbation at the turnaround point is 
$\Delta\rho_{\mathrm{T}} = 13.1$. 
For $\Omega_{\mathrm{m}} = 1$, 
the density fluctuation at the turnaround point is 
$\Delta\rho_{\mathrm{T}} = (3\pi/4)^{2} = 5.55$ 
(the density contrast $\delta_{\mathrm{T}} = \Delta\rho_{\mathrm{T}} - 1 = 4.55)$
\citep[see also ][]{2002sgd..book.....M}.

From Eq.~(\ref{eq:sph}) we can estimate the mass of a structure
at the turnaround point:
\begin{equation}
M_{\mathrm{T}}(R) = 1.16\times 10^{12}~\Omega_{\mathrm{m}}\Delta\rho_{\mathrm{T}}~
(R/h^{-1}~{\mathrm{Mpc}})^{3}h^{-1}M_\odot, 
\label{eq:mass1}
\end{equation} 
which gives us
\begin{equation}
M_{\mathrm{T}}(R) = 4.1\times 10^{12}~(R/h^{-1}~{\mathrm{Mpc}})^{3}h^{-1}M_\odot.
\label{eq:mass2}
\end{equation} 
The mass $M_{\mathrm{T}}(R)$ in the  Eq.~(\ref{eq:mass2}) 
describes the minimum mass needed in the sphere with radius $R$ 
for the turnaround and collapse \citep{2002MNRAS.337.1417G}.
We note that Planck results presently suggest $\Omega_{\mathrm{m}} = 0.308 \pm 0.012$
\citep{2015arXiv150201589P},
which in the flat model leads to $\Delta\rho_{\mathrm{T}} = 12.0$.

\begin{figure}[ht]
\centering
\resizebox{0.440\textwidth}{!}{\includegraphics[angle=0]{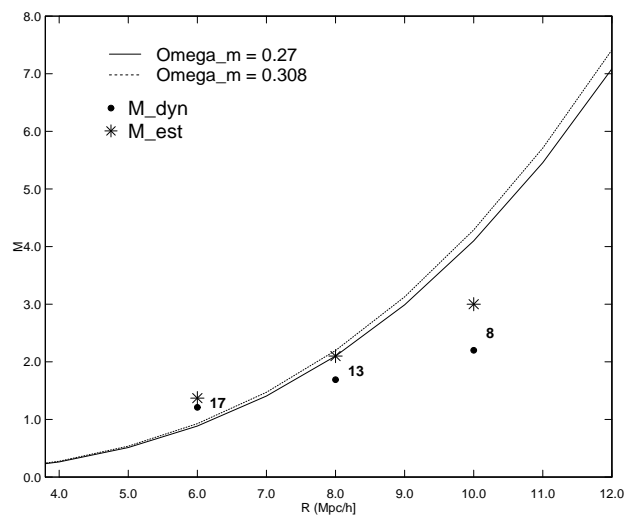}}
\caption{
Turnaround mass $M_{\mathrm{T}}(R)$ (in units of $10^{15}h^{-1}M_\odot$) 
versus radius of a sphere $R$  in a spherical
collapse model. The solid line corresponds to a flat model with
$\Omega_{\mathrm{m}} = 0.27$ and
grey dashed line to a  model with $\Omega_{\mathrm{m}} = 0.308$.
Filled circles show the total masses of galaxy groups in regions of different global 
density in the A2142 supercluster (Table~\ref{tab:D8prop}).   
Stars denote estimated masses as explained in the text.
The numbers show global density lower limits for each region.
}
\label{fig:mt}
\end{figure}

We can use  Eqs.~(\ref{eq:fluc})  and ~(\ref{eq:mass2})  
to examine the dynamical state of the A2142 supercluster. 
Table~\ref{tab:D8prop} summarises the properties of the different global 
luminosity density regions in  the A2142 supercluster. 
In Fig.~\ref{fig:mt} we plot turnaround mass $M_{\mathrm{T}}(R)$ 
versus radius of a sphere $R$
for $\Omega_{\mathrm{m}} = 0.27$ and for  $\Omega_{\mathrm{m}} = 0.308$,
and show in the figure both the dynamical and estimated
masses of different global density regions in the A2142 supercluster.

In the highest density core of the A2142 supercluster with $D8 \geq 17$ 
the sum of masses of galaxy groups is 
$M_{\mathrm{dyn}} = 1.2\times~10^{15}h^{-1}M_\odot$,
the estimated mass is approximately 10\% larger (Table~\ref{tab:D8prop}). 
The radius of this region is $R \approx 6$~\Mpc.
From Eq.~(\ref{eq:fluc}) we find that the density perturbation 
in the high-density core of A2142 supercluster is $\Delta\rho = 17.85$.
For $R = 6$~\Mpc\ the minimum mass needed for the turnaround is 
$M_{\mathrm{T}} = 0.89\times~10^{15}h^{-1}M_\odot$. 
This suggests that the core region in the A2142 supercluster 
has already reached the turnaround radius and started to collapse.

The radius of the region with $D8 \geq 13$ is $R \approx 8$~\Mpc. 
For this radius, the minimum mass for the turnaround is 
$M_{\mathrm{T}} = 2.1\times~10^{15}h^{-1}M_\odot$. 
The sum of dynamical masses of galaxy groups in this region is 
$M_{\mathrm{dyn}} = 1.7\times~10^{15}h^{-1}M_\odot$, 
which corresponds to $\Delta\rho = 10.5$.
This indicates that this region in the A2142 supercluster 
is very close to the turnaround state. 
If we add the estimated mass of faint groups 
and the estimated mass of intracluster gas in this region 
then the mass of the region with
global density $D8 \geq 13$  is $M_{\mathrm{est}} = 2.1\times~10^{15}h^{-1}M_\odot$,
sufficient for this region for the collapse (Table~\ref{tab:D8prop}). 
Therefore we cannot exclude the possibility that 
the mass of the whole $D8 \geq 13$ region
in A2142 supercluster is higher than the minimum mass needed for collapse. 
The turnaround radius in the A2142 supercluster is $R_T \approx 7-8$~\Mpc.
Further investigation of the properties of galaxy groups (and filaments)
in this region will help to understand the dynamical state of the region.

The size of the larger region with $D8 \geq 8$ is $R \approx 10$~\Mpc. 
The sum of dynamical masses of galaxy groups in this region is 
$M_{\mathrm{dyn}} = 2.2\times~10^{15}h^{-1}M_\odot$ which corresponds to 
$\Delta\rho = 7.1$. From Fig.~\ref{fig:mt} we also see that the estimated mass of this
region is not sufficient for this region to collapse. 
This shows that this region in the 
A2142 supercluster expands together with the universe.

We can also find masses and radii of two regions in the lower density
straight tail of  the A2142 supercluster around groups Gr14283 (12), and 
Gr10224 (13) and Gr26895 (14). The diameter of both regions is approximately 
$5$~\Mpc: there are seven poor groups in the region of Gr14283, and 
ten poor groups near groups Gr10224 and Gr26895. The sums of their
dynamical masses are $M_{\mathrm{dyn}} = 0.10\times~10^{15}h^{-1}M_\odot$
in the Gr14283 region
(estimated mass in this region 
$M_{\mathrm{est}} = 0.18\times~10^{15}h^{-1}M_\odot$), and 
$M_{\mathrm{dyn}} = 0.30\times~10^{15}h^{-1}M_\odot$ 
($M_{\mathrm{est}} = 0.42\times~10^{15}h^{-1}M_\odot$)
in the Gr10224 and Gr26895 region,
enough to become bound structures in the future.

\citet{2006MNRAS.366..803D}, 
\citet{2009MNRAS.399...97A}, and \citet{2011MNRAS.415..964L} 
studied the future evolution of superclusters
from observations and  simulations and showed that 
superclusters defined above a certain density contrast 
may become virialised systems in the future.
Such  superclusters become more 
spherical, and clusters in superclusters may merge into one cluster.
Density contrasts for future virialised systems for different cosmological models 
were calculated in \citet{2015A&A...575L..14C} who showed that
for $\Omega_{\mathrm{m}} = 0.27$ the density contrast limit is
$\Delta\rho = 8.73$.  
Comparison with the density contrasts in regions with different global
density limits shows that in region with $D8 \geq 8$
the density contrast in this region  is $\Delta\rho = 8.64$ 
if we use the value of the estimated mass of this region.   
Therefore, the A2142 supercluster 
may split into three systems in the future. 
The region with density contrast slightly higher than $D8 \geq 8$
will form one system, and  the regions around
the group Gr14283 and around the groups 
Gr10224 and Gr26895 will form two separate systems. 

\subsection{The shape of superclusters as a test for cosmological models}
\label{sect:shape} 

Recently \citet{2013ApJ...777...74S} and \citet{2014ApJ...784...84S}
analysed the straightness of superclusters in the $\Lambda$CDM model, and
in four different coupled dark energy models which
assume the existence of dark sector coupling between scalar field dark 
energy and non-baryonic dark matter, and in models which
adopted modified $f(R)$ gravity. They found that in coupled
dark energy models supercluster spines are less straight than in the
$\Lambda$CDM model, and $f(R)$ gravity also affects the shape
of superclusters. However, in their analysis the mean length of supercluster
spines was less than $2$~\Mpc, which is comparable to the size 
of the smallest galaxy filaments in  the A2142 supercluster. 
So it is not clear to which
objects from observations their superclusters correspond.

\citet{2002MNRAS.331.1020K}, \citet{2007A&A...462..397E} 
and \citet{2011MNRAS.411.1716C} proposed using
several shape parameters of superclusters as  tests for cosmological
models, among which the overall shape parameter defined using Minkowski
functionals and shapefinders. 
\citet{2007A&A...462..397E} compared 
the asymmetry of superclusters from 2dFGRS catalogue with
that of superclusters from the Millennium Run simulations
as characterised by the offset of the geometrical
mean centre from the dynamical one. They showed
that asymmetry of  simulated superclusters
is very close to that of real superclusters.
\citet{2011MNRAS.411.1716C} found that many superclusters are rather planar,
although among rich superclusters elongated, prolate structures dominate.
\citet{2011A&A...532A...5E}
calculated shape parameters for a large number of superclusters.
In their list, by overall shape according to this
parameter  the A2142 supercluster is elongated but not among
the most elongated superclusters. However, the overall shape parameter
does not characterise the straightness of superclusters well.

\citet{2007A&A...476..697E} showed that  the morphology of the richest 
superclusters from the SDSS DR7 catalogue and Millenium simulations is different. They found
among the richest superclusters from simulations some very planar systems,
while very planar systems are absent
among the observed superclusters, and no very dense and straight filament-like
systems. However, in this study
the number of superclusters under study was too small for statistical conclusions. 

\citet{2012ApJ...759L...7P} identified over- and under-dense large-scale structures from
the SDSS DR7 catalogue. They found that the most massive structures are typically
filamentary and are often looping. The richest over-dense
structure in the SDSS region was the dense part of the Sloan Great Wall, which looks like a
thick strip of galaxies with a length of about $150$~\Mpc. On the other hand,
the large under-dense structures are complexes of voids connected by tunnels. 
\citet{2012ApJ...759L...7P} also made quantitative comparisons 
between the observation and the mock SDSS
surveys performed in the simulated $\Lambda$CDM universe, using the Horizon Run 2
simulation of structure formation in a box with a side length of
$7200$~\Mpc\ \citep{2011JKAS...44..217K}. 
They found that the largest over-dense  structures from the
$\Lambda$CDM model frequently show massive filamentary structures similar to the
Sloan Great Wall. In particular, straight filaments are occasionally found. For
example, the second panel of Fig.~2 in \citet{2012ApJ...759L...7P} 
shows a straight filament with
a length of about $70$~\Mpc. It demonstrates that large-volume simulations are
essential in order to make quantitative comparisons of large-scale structures
between observations and cosmological models.

\subsection{Clusters, superclusters, and the cosmic web}
\label{sect:clenv}

\citet{1978IAUS...79..241J} were the first to note that rich galaxy clusters
are elongated, with 
axes oriented along the supercluster axis to which they belong.
The orientation of galaxy clusters along the supercluster
axis was later discussed in \citet{1980MNRAS.193..353E}, in \citet{1983ARA&A..21..373O},
and in \citet{1984ApJ...279....1D}. These early studies showed that the 
formation and evolution of galaxy clusters and their large-scale
environment is related; the alignment of clusters along the supercluster
axis reflects the existence of elongated superclusters that form the cosmic web
(cell structure, as the structure  of the Universe was called in these studies). 
The alignment of the galaxy groups and clusters
with the surrounding 
environments up to distances of about $30$~\Mpc\ 
has been detected both from observations and simulations 
\citep[][and references therein]{2005ApJ...618....1H, 2011MNRAS.414.2029P}. 
Recently, \citet{2015A&A...576L...5T} demonstrated that galaxy pairs are aligned
along the large-scale filaments where they are located.

\begin{figure}[ht]
\centering
\resizebox{0.225\textwidth}{!}{\includegraphics[angle=0]{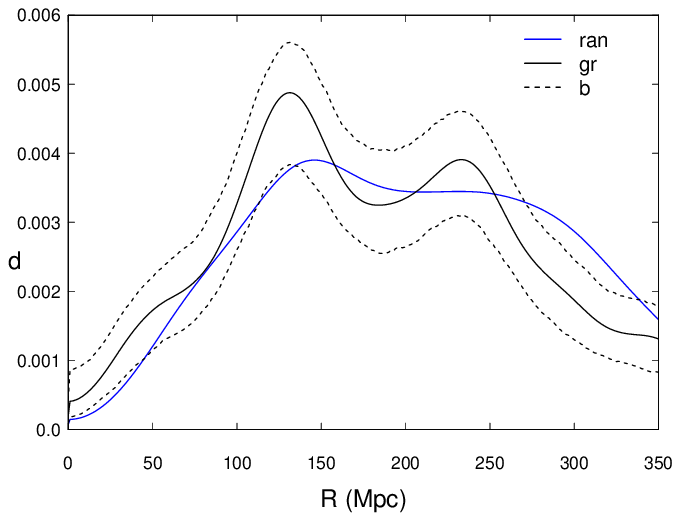}}
\resizebox{0.25\textwidth}{!}{\includegraphics[angle=0]{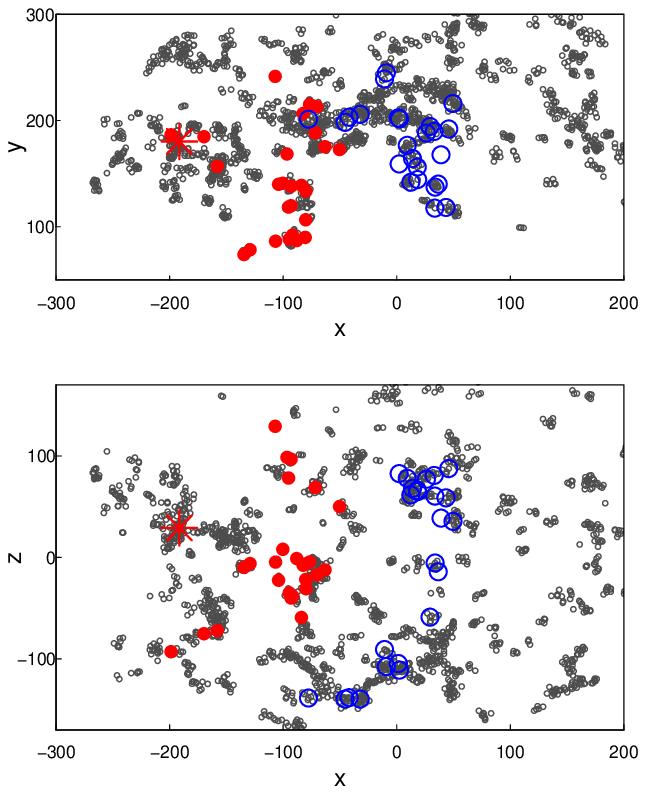}}
\caption{
Distribution of distances between  cluster A2142 and other
groups and clusters with 30  galaxies (left panel),
and the distribution of groups with at least
four galaxies in x, y, and z coordinates
(right panels). Gr denotes group
distributions, 'ran'   the random distribution, 
and 'b' the upper and lower 95\% 
bootstrap confidence limits for the group distributions.
In the right panels star shows the location of the cluster A2142,
red filled circles denote groups and clusters with at least
30 member galaxies forming the first maximum in the distance
distribution, and blue empty circles denote groups
and clusters that form the second maximum in the distance
distribution. Grey symbols mark the location of other groups
and clusters with at least four member galaxies in the region.
In the lower right panel the rich system at $-100 > x$~\Mpc\ 
and $z < -100$~\Mpc\ is the Sloan Great Wall.
} 
\label{fig:gr29587}
\end{figure}

This motivates us to analyse the distribution of other rich galaxy clusters
near A2142. We plot the distribution of distances to rich clusters
with at least 30 member galaxies from A2142 cluster, and the distribution
of galaxy groups and clusters in cartesian coordinates 
in the region around the A2142 supercluster up to $340$~\Mpc\ where the selections
are the smallest \citep{2011A&A...532A...5E}
in Fig.~\ref{fig:gr29587}. In the left panel we
 also show  bootstrap errors
of distance distributions, and the comparison of the 
distance distributions with distances to
a random sample of points found taking into account SDSS mask.
Interestingly, we see in the distance distribution two maxima, at $120-130$~\Mpc,
and at $240$~\Mpc. The first maximum is formed by galaxy groups and 
clusters in the superclusters across the void between the A2142 supercluster
and nearby superclusters, the Bootes supercluster among them.
Galaxy groups in the Sloan Great Wall and in other rich superclusters
cause a maximum at distances $D \approx 240$~\Mpc. 
Early studies of superclusters and of the supercluster-void network
have already demonstrated the presence of a characteristic scale of
about 120 - 130 \Mpc\ in the
distribution of rich superclusters
\citep{1994MNRAS.269..301E, 1997Natur.385..139E, 1997MNRAS.289..801E}. 
Recently  \citet{2012ApJ...749...81H} detected this scale
in the correlation function of a large sample of galaxy clusters. 

The scale of the first maximum is close to the well-known scale of the
baryonic acoustic oscillations (BAO). Planck results of the CMB
anisotropies determine the scale of acoustic oscillations -- 
$109$~$h^{-1}$~\Mpc\ with high accuracy \citep{2013arXiv1303.5076P}.
Could it be that the distribution of galaxy groups and clusters 
plotted in Fig.~\ref{fig:gr29587} mark the location of the local
BAO shells, as described in \citet{2012A&A...542A..34A}?

The location of the maximum differs from the location of 
the BAO scale. The maximum in
the distributions of distances of groups and clusters 
is wide, so it is not yet clear whether our results are in tension
with the strictly fixed BAO scale. 
Furthermore, BAO shells are barely visible in the distribution of
galaxies and their density contrast is rather low, while
in our case the clusters that cause the distance maxima are
located in rich superclusters, the richest galaxy system in the local Universe,
the Sloan Great Wall among them. The A2142 supercluster itself
marks the highest luminosity density value in the whole SDSS area.
This suggests that we might be dealing with a different phenomenon.
In a forthcoming paper we will search for characteristic scales in
the distribution of galaxy groups and clusters in more detail.

\citet{2011A&A...531A..75E, 2011A&A...534A.128E} and 
\citet{2011A&A...531A.149S} showed how the density 
waves of different scales affect the richness of galaxy systems. 
Rich galaxy clusters and high-density cores of superclusters 
form in regions of high environmental density, 
where positive sections of medium- and large-scale density 
perturbations combine. The most luminous groups and clusters from 
observations and simulations are located in high-density regions, often 
in cores of rich superclusters 
\citep{2003A&A...401..851E, 2005A&A...436...17E, 2012A&A...542A..36E},
which are ideal places to study cluster
merging \citep{2002AJ....123.1216R, 2004ogci.conf...71B}.
Matter flows to regions of deep potential wells where 
the richness and sizes of galaxy groups and clusters increase via
merging. Syncronisation of the orientations of galaxy clusters
and supercluster axes occurs as a result of these processes.
The A2142 supercluster with its collapsing core and multimodal galaxy groups and
clusters is a very good example of these multiscale
processes  both in the high-density core and low-density outskirts.

\subsection{Summary}
\label{sect:sum} 

Summarising, our results are as follows.

\begin{itemize}
\item[1)]
Rich galaxy groups and clusters in the A2142 supercluster lie along an almost
straight line. Morphologically A2142 supercluster resembles a rich and straight
filament with a high-density spider-like core.
\item[2)]
Most of the rich groups in the supercluster are concentrated in the high-density core 
region with $D8 > 13$.
The outskirts of the supercluster is populated mostly by poor groups.
\item[3)]
The fraction of stellar
mass in  the A2142 supercluster is about $1-2$\% of the total mass, being slightly higher
in the outskirts of the supercluster.
\item[4)]
In the high-density core seven out of eight rich galaxy groups with more than twenty
galaxies are multimodal. 
\item[5)]
The richest cluster of the supercluster, Abell cluster A2142 has an elongated shape; 
the orientation of the cluster axis follows 
the orientations of its X-ray substructures and radio halo  
and it is aligned along the supercluster axis. 
\item[6)] 
The high-density core region in the A2142 supercluster with global density
$D8 \geq 17$ and probably also with $D8 \geq 13$ has
reached the turnaround radius and started to collapse. 
\item[7)] 
The A2142 supercluster with its high-density, collapsing core and 
long filament-like
tail with two density enhancements 
may split into several systems in the future. 

\end{itemize}

The A2142 supercluster with its collapsing  core and straight body,
in which the richest clusters are aligned along the supercluster axis,
is an unusual object among galaxy superclusters.
We plan to continue our studies of the A2142 supercluster with the analysis of the galaxy
content of the supercluster. We also will search for possible collapsing 
high-density cores in other superclusters, and will study the morphological
properties and search for collapsing superclusters from simulations.
Such studies help us to understand the role of the dark matter and dark energy 
in the formation and evolution of the galaxy superclusters,
the largest structures in the cosmic web.

\begin{acknowledgements}

We thank the referee for useful comments which helped to improve the paper.
We thank Prof. Peeter Tenjes for valuable discussions.
We are pleased to thank the SDSS Team for the publicly available data
releases.  Funding for the Sloan Digital Sky Survey (SDSS) and SDSS-II has been
provided by the Alfred P. Sloan Foundation, the Participating Institutions,
the National Science Foundation, the U.S.  Department of Energy, the
National Aeronautics and Space Administration, the Japanese Monbukagakusho,
and the Max Planck Society, and the Higher Education Funding Council for
England.  The SDSS Web site is \texttt{http://www.sdss.org/}.
The SDSS is managed by the Astrophysical Research Consortium (ARC) for the
Participating Institutions.  The Participating Institutions are the American
Museum of Natural History, Astrophysical Institute Potsdam, University of
Basel, University of Cambridge, Case Western Reserve University, The
University of Chicago, Drexel University, Fermilab, the Institute for
Advanced Study, the Japan Participation Group, The Johns Hopkins University,
the Joint Institute for Nuclear Astrophysics, the Kavli Institute for
Particle Astrophysics and Cosmology, the Korean Scientist Group, the Chinese
Academy of Sciences (LAMOST), Los Alamos National Laboratory, the
Max-Planck-Institute for Astronomy (MPIA), the Max-Planck-Institute for
Astrophysics (MPA), New Mexico State University, Ohio State University,
University of Pittsburgh, University of Portsmouth, Princeton University,
the United States Naval Observatory, and the University of Washington.

The present study was supported by the ETAG projects 
IUT26-2 and IUT40-2, and by the European Structural Funds
grant for the Centre of Excellence "Dark Matter in (Astro)particle Physics and
Cosmology" TK120. JN acknowledges ETAG grant PUT246.
This work has also been supported by
ICRAnet through a professorship for Jaan Einasto.

\end{acknowledgements}

\bibliographystyle{aa}
\bibliography{scl001.bib}

\end{document}